\documentclass[runningheads]{llncs}
\usepackage[T1]{fontenc}
\usepackage{amsmath,amssymb}
\usepackage{stmaryrd}   
\usepackage{graphicx}
\usepackage{booktabs}
\usepackage{listings}
\usepackage{xcolor}
\usepackage{xspace}
\usepackage{tikz}
\usepackage{cite}
\usetikzlibrary{arrows.meta,backgrounds,fit}
\usepackage{pgfplots}
\pgfplotsset{compat=1.18}
\usepackage{hyperref}

\usepackage[htt]{hyphenat} 
\def\orcidID#1{\smash{\href{http://orcid.org/#1}{\protect\raisebox{-1.25pt}{\protect\includegraphics{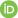}}}}}

\usepackage{todonotes}

\lstdefinelanguage{lean}{
  morekeywords={theorem,def,lemma,import,namespace,end,by,open,structure,inductive,where,let,fun,match,with},
  morecomment=[l]{--},
  morecomment=[s]{/-}{-/},
  sensitive=true
}
\newcommand{\sem}[2]{\llbracket #1\rrbracket_{#2}}  
\newcommand{\val}[2]{\beta_{#1,#2}}                 
\newcommand{\Cr}{C'}                                

\newcommand{\amulettwo}{\textsc{AMulet2}\xspace}
\newcommand{\talismantwo}{\textsc{TalisMan2}\xspace}
\newcommand{\pacheck}{\textsc{Pacheck}\xspace}
\newcommand{\pasteque}{\textsc{Past\`eque}\xspace}
\newcommand{\addersub}{\texttt{Adder Substitution}\xspace}
\newcommand{\Addersub}{\texttt{Adder Substitution}\xspace}
\newcommand{\guessprove}{\texttt{Guess-and-Prove}\xspace}
\definecolor{amcol}{HTML}{0072B2}  
\definecolor{tmcol}{HTML}{D55E00}  

\begin{document}
\title{Composing SAT and Algebraic Certificates in~Lean for End-to-End Circuit Verification\thanks{This work was partially 
supported by 
the Austrian Science Fund (FWF) [10.55776/ESP666];
a gift from Amazon Web Services; and
Coordenação de Aperfeiçoamento de Pessoal de Nível Superior - Brasil (CAPES) -
Finance Code 001;}}
\titlerunning{Composing SAT and Algebraic Certificates}
\author{Daniela Kaufmann\inst{1}\orcidID{0000-0002-5645-0292} \and   Pedro Saccomani\inst{2}\orcidID{0009-0007-2854-0104}}
%
\authorrunning{D. Kaufmann, P. Saccomani}
%
\institute{TU Wien, Austria \and   Universidade Federal de Minas Gerais, Brazil}
\maketitle              
\setcounter{footnote}{0}
\begin{abstract}
State-of-the-art tools for the verification of multiplier circuits combine
computer algebra with SAT solving. \amulettwo replaces a final-stage adder that
algebraic reasoning cannot handle efficiently and uses SAT to justify this
replacement. \talismantwo uses SAT to prove guessed linear relations, which serve
as additional algebraic axioms. 
Proof logging increases the trust in both tools. 
The algebraic part can be
certified by a PAC certificate and the SAT part by LRAT refutations. Checking
both certificates separately, however, does not establish that the circuit is
correct, because no existing checker relates the axioms of the PAC
certificate to the input circuit or to the facts established by the SAT solver. In
this paper, we close this gap by composing both certificates into a single Lean~4 theorem stating
that the circuit in the input AIGER file is a correct multiplier. 
We show that
each method needs SAT for exactly one step of this proof and discharge this
step with LRAT refutations. Both methods certify all 192 AOKI multipliers with 64-bit operands and
Lean checks each theorem in under seven minutes.
\keywords{Certified Verification \and Lean \and Computer Algebra
  \and Circuit Verification \and Proof Certificates}
\end{abstract}

\section{Introduction}
\label{sec:intro}

Current state-of-the-art techniques for the verification of multiplier
circuits combine word-level reasoning based on computer algebra with bit-level
reasoning based on SAT solving. Algebraic methods model each gate in the circuit by a
polynomial and prove the word-level specification by reducing the
specification polynomial by the gate
polynomials~\cite{CiesielskiEtAl-DAC15,SayedAhmedEtAl-DATE16,RitircBiereKauers-FMCAD17,KaufmannBiereKauers-FMCAD19}. For
certain multiplier architectures, this reduction is efficient for operand widths
of 64 bits and
beyond~\cite{MahzoonGrosseDrechsler-ICCAD18,MahzoonGrosseDrechsler-DAC19,KonradScholl-FMCAD24,KaufmannBerthomieu-TACAS25}.
The reduction can be recorded as a certificate in the Practical Algebraic Calculus
(PAC)~\cite{RitircBiereKauers-SCSC18,KaufmannFleuryBiere-FMCAD20,KaufmannFleuryBiereKauers-PAC-FMSD22},
which can be checked independently.

However, many architectures cannot be verified efficiently by algebraic
reasoning alone. If the final-stage adder in a multiplier is a carry-lookahead or
parallel-prefix adder, the intermediate reduction polynomials
blow up~\cite{MahzoonGrosseDrechsler-DAC19,KaufmannBiereKauers-FMCAD19}. Two recent tools therefore include
SAT solving to simplify the polynomial encoding. \Addersub~\cite{KaufmannBiereKauers-FMCAD19}, implemented in
\amulettwo~\cite{KaufmannBiere-TACAS21,KaufmannBiere-STTT23}, replaces the
complex final-stage adder by a simple ripple-carry adder, checks the equivalence
of the rewriting using SAT, and verifies the rewritten circuit algebraically.
\guessprove~\cite{HofstadlerKaufmann-CP25,HofstadlerKaufmannChen-2026},
implemented in \talismantwo~\cite{TalismanTool} guesses linear relations between
circuit gates and proves their correctness using SAT. These relations are used as
additional axioms during the algebraic reduction of the specification, which has
additionally been linearized using extension variables. 
This allows to keep the whole algebraic rewriting in \talismantwo linear.

Since verification tools may contain bugs, proof logging is used to increase
trust in their results. For both tools, each part of the proof can be certified
separately, the algebraic part by a PAC certificate and the SAT part by an LRAT
refutation~\cite{CruzFilipeEtAl-RAT-CADE17}, and verified checkers exist for both kinds of
certificates~\cite{KaufmannFleuryBiere-FMCAD20,Lammich-GRAT-JAR20,TanHeuleMyreen-CakeLPR-TACAS21}.

However, even if all certificates are accepted, the correctness of the circuit
is not fully established. First, the PAC checker receives the gate
polynomials from \amulettwo or \talismantwo, which derive them from the input
And-Inverter Graph (AIG). If this translation is incorrect, the checker establishes a correct
statement about a different circuit. Second, the results of the two checkers
are not linked. An LRAT checker establishes that a CNF is unsatisfiable, but
not what this CNF expresses, and nothing ensures that it expresses precisely
the fact that the algebraic proof assumes. If the CNF in \amulettwo compared
the wrong pair of circuits, or if the CNF in \talismantwo encoded a guessed
relation that differs from the corresponding axiom of the certificate, both
certificates would be accepted, although the combined argument is invalid.
In principle, the SAT proof can be translated into PAC to obtain a single
certificate, but the overhead makes this infeasible in
practice~\cite{KaufmannBiereKauers-DATE20}. And this would still not solve the first issue.

\paragraph{Contributions.} We compose the PAC certificate and the LRAT
refutations into one Lean~4~\cite{deMouraUllrich-CADE21} theorem stating that
the multiplier circuit in the input AIGER~\cite{Biere-AIGER07} file computes the product
of its operands:
\begin{itemize}
  \item We
        reduce this theorem to five proof obligations (Sect.~\ref{sec:recipe}).
        One is the validity of the PAC certificate, and the others connect the
        certificate to the input circuit and to the integer specification.
        \Addersub and \guessprove discharges exactly one of them by SAT.
\item We obtain this SAT-justified obligation from LRAT refutations through a generic bridge, which utilizes the verified LRAT checker of \texttt{bv\_decide}~\cite{BovingEtAl-BVDecide-OOPSLA25} without changes~(Sect.~\ref{sec:bridge}).
\item We instantiate the bridge for both methods (Sect.~\ref{sec:instances}).
  For \addersub, it shows that the rewritten circuit is equivalent
  to the input circuit, and for \guessprove, that the guessed relations
  hold.
\item We implement a reflective PAC checker and an AIGER parser in Lean,
  which both methods share (Sect.~\ref{sec:impl}), and certify all 192
  multipliers with 64-bit operands of the AOKI benchmarks with both methods
  (Sect.~\ref{sec:eval}).
\end{itemize}

All lemmas, including the soundness of the checkers and of the bridge, are
checked by the Lean kernel. Only the final step, the evaluation of the
combined check on the data in the generated file, is carried out by compiled
code via \texttt{native\_decide}, because evaluation by the kernel is
infeasible at this scale. Besides the kernel, we therefore trust the Lean
compiler for this single evaluation and the statement of the final theorem~(Sects.~\ref{sec:transcribe} and~\ref{sec:trust}).

\section{Preliminaries}
\label{sec:bg}

\paragraph{And-Inverter Graphs.}
An AIG~\cite{KuehlmannEtAl-RobustEC-TCAD02} is a directed acyclic graph
representing a Boolean circuit~$C$. Internal nodes are conjunctions of two
inputs, and edge annotations encode negation. We call the inputs, gates and
outputs of $C$ the variables of~$C$. A literal is a variable, its negation
or a constant. 

A circuit $C$ is \emph{well formed} if its gates are topologically
ordered, that is, the inputs of every gate are constants, primary inputs or
earlier gates, and every output refers to an existing variable. For an input
assignment~$i$, which maps each primary input of~$C$ to $0$ or~$1$,
evaluating the gates in order yields a Boolean value
$\sem{C}{i}(\ell)$ for every literal~$\ell$.

\begin{definition}[Multiplier]
\label{def:mult}
Let $C$ be a well-formed circuit whose inputs are the operand bits
$a_0,\dots,a_{n-1}$ and $b_0,\dots,b_{n-1}$ and whose outputs are
$s_0,\dots,s_{2n-1}$. Then $C$ is a \emph{multiplier} if $S_C(i) = A_C(i)\cdot B_C(i)$ for every input
assignment~$i$, with $A_C(i) = \sum_{k=0}^{n-1} 2^k \sem{C}{i}(a_k)$,
$B_C(i) = \sum_{k=0}^{n-1} 2^k \sem{C}{i}(b_k)$ and
$S_C(i) = \sum_{j=0}^{2n-1} 2^j \sem{C}{i}(s_j)$.
\end{definition}

\paragraph{Algebraic verification.}
Algebraic methods encode $C$ by polynomials in~$\mathbb{Z}[X]$, where $X$
denotes the variables of~$C$. Since every variable takes only the values $0$
and~$1$, it satisfies $x^2 = x$, which is expressed by the \emph{Boolean value
constraint} $x^2 - x$. Based on the correspondences
$x \wedge y \simeq xy$ and $\neg x \simeq 1-x$, each gate
$g = \ell_1 \wedge \ell_2$ yields the \emph{gate polynomial}
$-g + \hat\ell_1\hat\ell_2$, where $\hat\ell$ is $x$ or $1-x$ for a positive
or negative literal of~$x$ and $0$ or~$1$ for a constant. We denote by~$G(C)$
the set of these polynomials. The \emph{specification polynomial} is
$P_C = S - A\cdot B$ with $S = \sum_{j=0}^{2n-1} 2^j s_j$,
$A = \sum_{k=0}^{n-1} 2^k a_k$ and $B = \sum_{k=0}^{n-1} 2^k b_k$.

Since $S_C(i)$ and $A_C(i)\cdot B_C(i)$ both lie in $[0,2^{2n})$, it suffices
to show $S_C(i) \equiv A_C(i)\cdot B_C(i) \pmod{2^{2n}}$. We denote by
$\mathrm{Ax}(C) = G(C) \cup \{2^{2n}\}$ the axioms of~$C$, where the
constant~$2^{2n}$, called the \emph{modulus}, allows coefficients that are
multiples of~$2^{2n}$ to be discarded. Accordingly, polynomials are evaluated
in the ring $R = \mathbb{Z}/2^{2n}\mathbb{Z}$. A \emph{Boolean
valuation}~$\sigma$ maps every variable to $0$ or~$1$ in~$R$. It extends
uniquely to a ring homomorphism $\sigma : \mathbb{Z}[X] \to R$, under which
every Boolean value constraint vanishes. For an input assignment~$i$, the Boolean
valuation~$\val{C}{i}$ maps each variable to its value under~$i$. The circuit $C$ multiplies if $P_C$ lies in the ideal
generated by $\mathrm{Ax}(C)$ and the Boolean value constraints $x^2-x$. 
This ideal membership can be decided by reducing
$P_C$ by the gate polynomials from the outputs towards the
inputs~\cite{RitircBiereKauers-FMCAD17}.

\paragraph{Linearized specification.}
\guessprove~\cite{HofstadlerKaufmann-CP25,HofstadlerKaufmannChen-2026}
does not reduce~$P_C$ itself, but first
\emph{linearizes} it. It extends~$C$ by a gate $t_{kl} = a_k \wedge b_l$ for
every partial product and replaces $A\cdot B$ by a sum over these gates:
\[
P_{\mathrm{TM}} \;=\; \sum_{j=0}^{2n-1} 2^j\, s_j \;-\; \sum_{k=0}^{n-1}\sum_{l=0}^{n-1} 2^{k+l}\, t_{kl}.
\]
Since $t_{kl}$ evaluates to $a_k b_l$, both polynomials agree under every
input assignment.

\paragraph{PAC certificates.}
The polynomial reduction of the specification can be recorded as a PAC
certificate~\cite{KaufmannFleuryBiere-FMCAD20,KaufmannFleuryBiereKauers-PAC-FMSD22}.
The extension LPAC~\cite{KaufmannFleuryBiereKauers-PAC-FMSD22} provides a
linear-combination rule, which derives $\sum_{j=1}^{m} c_j\, p_j$ from
antecedents~$p_1,\dots,p_m$ and arbitrary cofactors~$c_1,\dots,c_m$ in a
single step, and
\amulettwo emits one such step for each substitution during the reduction.
We use the following fragment of LPAC, in which the Boolean value
constraints are built in by normalizing monomials with respect to $x^2 = x$:

\begin{definition}[PAC certificate]
\label{def:cert}
A \emph{PAC certificate} $\Pi$ consists of a finite set $A(\Pi)$ of
\emph{axioms}, a sequence of steps and a \emph{target} $t(\Pi)$, all
in~$\mathbb{Z}[X]$. Step~$k$ states a \emph{conclusion} $q_k$ together with a
representation $q_k = \sum_{j=1}^{m_k} c_{kj}\, r_{kj}$, where each $r_{kj}$ is an
axiom or an earlier conclusion and the cofactors $c_{kj} \in \mathbb{Z}[X]$
are arbitrary. $\Pi$ is \emph{valid} if every stated representation holds
modulo $x^2 = x$ and the last conclusion equals $t(\Pi)$ modulo $x^2 = x$.
\end{definition}

\noindent
A valid certificate thus shows that $t(\Pi)$ lies in the ideal generated
by~$A(\Pi)$ and the Boolean value constraints.
For a multiplier, the axioms are~$\mathrm{Ax}(C)$ and the target is~$P_C$.
A checker decides validity by recomputing every sum, which requires no
search. Existing checkers are \pacheck, an efficient
implementation in~C, and \pasteque, which is verified in
Isabelle/HOL~\cite{KaufmannFleuryBiere-FMCAD20}.

\paragraph{SAT and LRAT.}
A SAT solver decides whether a formula in conjunctive normal form (CNF), a
conjunction of clauses, has a satisfying assignment.
The Tseitin encoding~\cite{Tseitin-1983} of an AIG uses one CNF variable per
AIG variable and, for each gate $g = \ell_1\wedge\ell_2$, the clauses
$(\neg g \vee \ell_1)$, $(\neg g \vee \ell_2)$ and
$(g \vee \neg\ell_1 \vee \neg\ell_2)$. Every input assignment~$i$ extends
uniquely to a satisfying assignment of these clauses, namely~$\sem{C}{i}$. A
\emph{miter}~\cite{GoldbergPradsadBrayton-DATE01} proves two circuits
equivalent by connecting them at their inputs, adding clauses asserting that
some pair of corresponding outputs differs, and showing that the resulting
CNF is unsatisfiable.

To certify unsatisfiability, SAT solvers such as
CaDiCaL~\cite{PollittEtAl-CaDiCaL3-SAT26} can emit an LRAT
proof~\cite{CruzFilipeEtAl-RAT-CADE17}, consisting of clause additions and
deletions that eventually derive the empty clause. Each clause addition is
annotated with hints identifying the clauses used to justify the addition. These
hints enable a checker to validate the proof without search.

\paragraph{Lean and proof by reflection.}
Lean~4~\cite{deMouraUllrich-CADE21} is a theorem prover whose proofs are checked
by a small kernel. Properties of large concrete objects are proved by
\emph{reflection}: one proves once that a Boolean function~$f$ returns
\texttt{true} only on objects with the property, and obtains $f\,d =
\mathtt{true}$ for a concrete~$d$ by evaluation. The tactic \texttt{decide}
evaluates $f\,d$ by kernel reduction, which adds nothing to the trusted base but
is intended for small computations. The tactic \texttt{native\_decide} executes
compiled code instead of kernel reduction and justifies the result using an
auxiliary axiom, thereby relying on the correctness of the native evaluation
mechanism of Lean.

The tactic \texttt{bv\_decide}~\cite{BovingEtAl-BVDecide-OOPSLA25}
checks LRAT refutations with a verified function \texttt{verifyCert}, whose
correctness theorem \texttt{verifyCert\_correct} states that
$\texttt{verifyCert}\ \mathit{cnf}\ \mathit{cert} = \mathtt{true}$ implies that
$\mathit{cnf}$ is unsatisfiable.

\section{Composing SAT and Algebraic Proofs}
\label{sec:recipe}

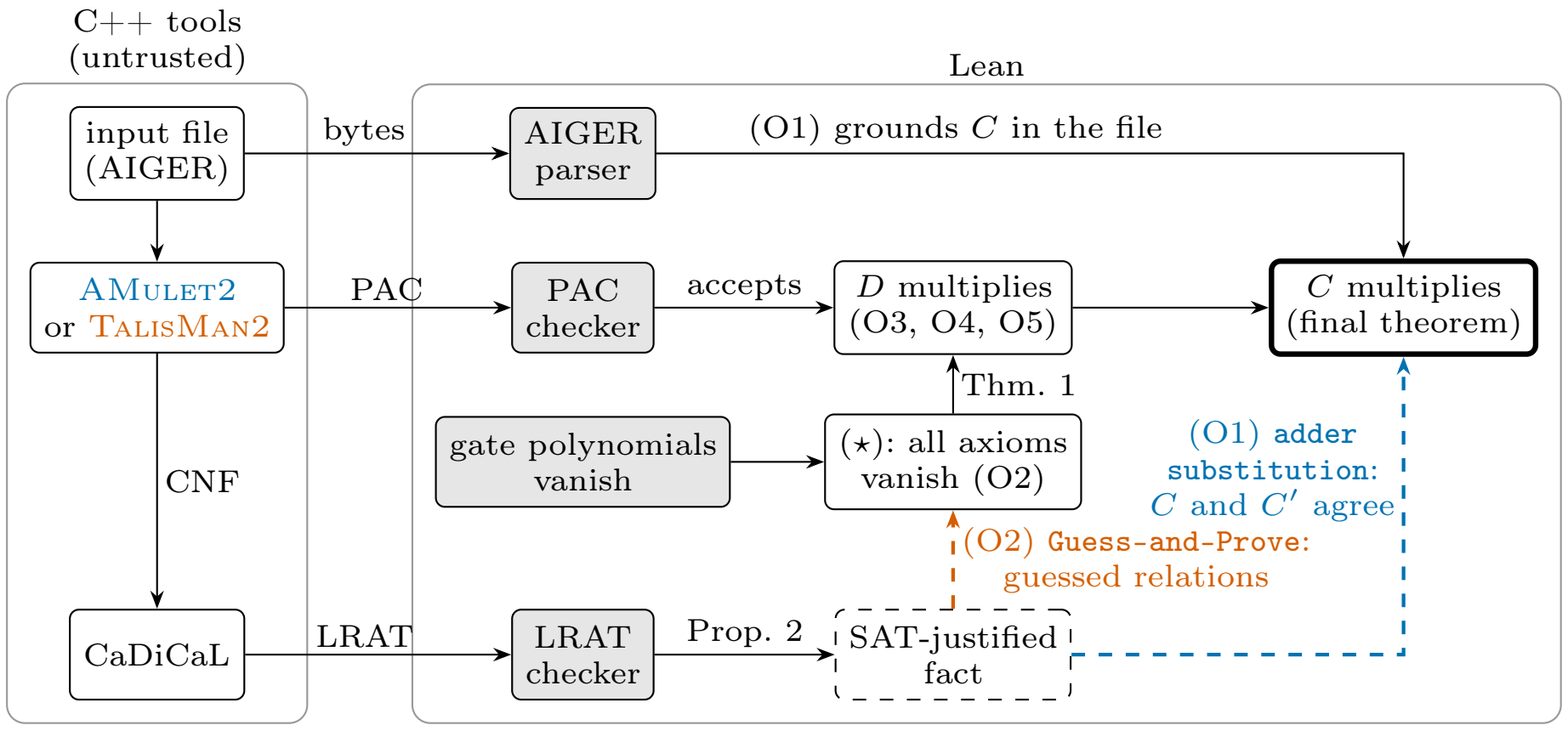
\begin{figure}[t]
\centering
\begin{tikzpicture}[
  font=\scriptsize,
  x=0.96cm,
  >={Stealth[length=4pt]},
  box/.style={draw,rounded corners=2pt,align=center,inner sep=3pt,minimum height=7mm},
  shared/.style={box,fill=black!10},
  satfact/.style={box,dashed},
  lab/.style={inner sep=1.5pt,align=center},
  region/.style={draw=black!40,rounded corners=4pt,inner sep=5pt},
]
\def\yA{0}\def\yB{-1.2}\def\yC{-2.4}\def\yD{-3.9}
\node[box]      (file) at (0.75,\yA) {input file\\(AIGER)};
\node[box]      (tool) at (0.75,\yB) {{\color{amcol}\amulettwo}\\or {\color{tmcol}\talismantwo}};
\node[box]      (sat)  at (0.75,\yD) {CaDiCaL};
\draw[->] (file) -- (tool);
\draw[->] (tool) -- node[lab,right] {CNF} (sat);
\node[shared]   (parse) at (4.2,\yA) {AIGER\\parser};
\node[shared]   (cc)    at (4.2,\yB) {PAC\\checker};
\node[shared]   (gates) at (4.2,\yC) {gate polynomials\\vanish};
\node[shared]   (vc)    at (4.2,\yD) {LRAT\\checker};
\draw[->] (file) -- node[lab,above,pos=0.45] {bytes} (parse);
\draw[->] (tool) -- node[lab,above,pos=0.45] {PAC} (cc);
\draw[->] (sat)  -- node[lab,above,pos=0.45] {LRAT} (vc);
\node[box]      (star)  at (7.2,\yC) {$(\star)$: all axioms\\vanish (O2)};
\node[box]      (alg)   at (7.2,\yB) {$D$ multiplies\\(O3, O4, O5)};
\node[satfact]  (fact)  at (7.2,\yD) {SAT-justified\\fact};
\node[box,very thick] (thm) at (10.85,\yB) {$C$ multiplies\\(final theorem)};
\draw[->] (gates) -- (star);
\draw[->] (cc)    -- node[lab,above] {accepts} (alg);
\draw[->] (star)  -- node[lab,right] {Thm.~\ref{thm:relsound}} (alg);
\draw[->] (vc)    -- node[lab,above] {Prop.~\ref{prop:bridge}} (fact);
\draw[->] (alg)   -- (thm);
\draw[->] (parse) -- node[lab,above,pos=0.4] {(O1) grounds $C$ in the file} (thm.north |- parse) -- (thm);
\draw[->,dashed,thick,tmcol] (fact) -- node[lab,right] {(O2) \guessprove:\\guessed relations} (star);
\draw[->,dashed,thick,amcol] (fact) -- (thm.south |- fact) -- node[lab,left,pos=0.62] {(O1) \texttt{adder}\\\texttt{substitution}:\\$C$ and $\Cr$ agree} (thm);
\begin{scope}[on background layer]
  \node[region,fit=(file)(sat)(tool),label={[lab]above:C++ tools\\(untrusted)}] {};
  \node[region,fit=(parse)(gates)(vc)(thm)(fact),label={[lab]above:Lean}] {};
\end{scope}
\end{tikzpicture}
\caption{Overview of the proof. Shaded components are proved or defined once
and shared by both methods. The SAT-justified fact discharges (O1) in
\addersub ({\color{amcol}blue}) and (O2) in \guessprove
({\color{tmcol}orange}).}
\label{fig:method}
\end{figure}

In this section, we combine a PAC certificate and LRAT refutations into a
single theorem stating that the circuit~$C$ in an AIGER file is a correct multiplier. We
reduce this theorem to five proof obligations (Sect.~\ref{sec:relsound}),
three of which tie the certificate to the input circuit~$C$. In purely
algebraic verification, which can occur in \amulettwo whenever the final-stage adder is not a parallel prefix adder,
these three are discharged by parsing the input
file in Lean and comparing the PAC certificate with the parsed circuit
(Sect.~\ref{sec:discharge}). In each SAT-assisted
method, \addersub and \guessprove, exactly one of the other two obligations depends on a SAT-justified fact instead
(Sect.~\ref{sec:twoplaces}), and we derive it from an LRAT refutation
(Sect.~\ref{sec:bridge}). Figure~\ref{fig:method} outlines the proof scheme.

We first state what a valid PAC certificate guarantees: its target
vanishes under every Boolean valuation under which all of its axioms vanish.

\begin{theorem}[Relative soundness]
\label{thm:relsound}
Let $\Pi$ be a valid PAC certificate and $\sigma$ a Boolean valuation. If
\[
(\star)\qquad \sigma(p) = 0 \quad\text{for every } p \in A(\Pi),
\]
then $\sigma(t(\Pi)) = 0$.
\end{theorem}

\begin{proof}
Since $\sigma$ is Boolean, $\sigma(x^2 - x) = 0$ for every variable~$x$, so
polynomials that are equal modulo $x^2 = x$ have the same value
under~$\sigma$. By induction on the steps, each conclusion $q_k$ is a
combination of polynomials that vanish under~$\sigma$, and hence
$\sigma(q_k) = 0$. The claim follows because $t(\Pi)$ equals the last
conclusion modulo $x^2 = x$.
\qed\end{proof}

Theorem~\ref{thm:relsound} does not establish the correctness of the axioms~$(\star)$. The existing PAC
checker \pacheck~\cite{KaufmannFleuryBiere-FMCAD20} provides the same
relative guarantee, namely that the target lies in the ideal generated by the
axioms. However, \pacheck does not verify that the axioms are correctly inferred from
the given circuit $C$.
For correctness of the axioms, we need the following lemma, saying that
 $(\star)$ holds under every circuit valuation:

\begin{lemma}[Circuit axioms]
\label{lem:circax}
Let $C$ be well formed. Then $\val{C}{i}(p) = 0$ for every
$p \in \mathrm{Ax}(C)$ and every input assignment~$i$.
\end{lemma}

\begin{proof}
On $\{0,1\}$, conjunction coincides with multiplication and negation with
$x \mapsto 1-x$. Hence every polynomial of~$G(C)$ vanishes under
$\val{C}{i}$; well-formedness ensures that every gate is evaluated from
values that are already defined.
\qed\end{proof}

\subsection{The Theorem and its Proof Obligations}
\label{sec:relsound}

By Theorem~\ref{thm:relsound} and Lemma~\ref{lem:circax}, a valid PAC
certificate~$\Pi$ with $A(\Pi) \subseteq \mathrm{Ax}(C)$ establishes
$\val{C}{i}(t(\Pi)) = 0$ for every input assignment~$i$. However, a PAC certificate consists of polynomials only
and does not contain a circuit. Its axioms are typically the gate polynomials
of a circuit~$D$, which need not be the circuit~$C$ of the input file:
\amulettwo generates the PAC certificate for a rewritten copy of~$C$,
and \talismantwo for an extension of~$C$ (Sect.~\ref{sec:twoplaces}). We
call $D$ the \emph{certificate's circuit}. To conclude that $C$ multiplies,
$D$ has to be related to~$C$, the axioms~$A(\Pi)$ to the circuit
axioms~$\mathrm{Ax}(D)$, and the target~$t(\Pi)$ to the
specification~$P_D$. The correctness of~$C$ thus reduces to the following
five proof obligations:

\begin{description}
\item[(O1) Circuit.] $C$ and $D$ are equivalent. That is, they have the same primary inputs, and under every
  input assignment their operand bits and their output bits have the same
  values.
\item[(O2) Axioms.] $\val{D}{i}(p) = 0$ for every $p \in A(\Pi)$ and every
  input assignment~$i$.
\item[(O3) Validity.] The certificate $\Pi$ is valid.
\item[(O4) Target.] $\val{D}{i}(t(\Pi)) = \val{D}{i}(P_D)$ for every input
  assignment~$i$.
\item[(O5) Lifting.] For every input assignment~$i$, if $\val{D}{i}(P_D) = 0$,
  then $S_D(i) = A_D(i)\cdot B_D(i)$.
\end{description}

\begin{proposition}[Decomposition]
\label{prop:decomp}
Let $C$ and $D$ be well-formed circuits, and let $\Pi$ be a PAC certificate. If (O1)--(O5) hold, then $C$ multiplies.
\end{proposition}

\begin{proof}
Let $i$ be an input assignment. By (O2) and (O3), Theorem~\ref{thm:relsound}
applies to $\sigma = \val{D}{i}$ and yields $\val{D}{i}(t(\Pi)) = 0$. By (O4),
$\val{D}{i}(P_D) = 0$, and (O5) gives $S_D(i) = A_D(i)\cdot B_D(i)$. By (O1),
$C$ and~$D$ have the same operand and output bits under~$i$, so
$A_C(i) = A_D(i)$, $B_C(i) = B_D(i)$ and $S_C(i) = S_D(i)$. Hence
$S_C(i) = A_C(i)\cdot B_C(i)$.
\qed\end{proof}

Since the PAC certificate may use the modulus $2^{2n}$ 
to reduce large coefficients, it establishes the specification
only modulo~$2^{2n}$. Obligation (O5) lifts this congruence to an equality
of integers. Since $S_D(i)$ and $A_D(i)\cdot B_D(i)$ lie in $[0,2^{2n})$, it
holds for every well-formed circuit:

\begin{lemma}[Lifting]
\label{lem:lift}
Let $D$ be well formed and let $i$ be an input assignment. If
$\val{D}{i}(P_D) = 0$, then $S_D(i) = A_D(i)\cdot B_D(i)$.
\end{lemma}

\begin{proof}
Let $i$ be an input assignment. Since $\val{D}{i}$ maps every variable to $0$
or~$1$, $\val{D}{i}(P_D)$ is the residue modulo $2^{2n}$ of the integer
\[
\delta \;=\; S_D(i) - A_D(i)\cdot B_D(i).
\]
The output has $2n$ bits, so $0 \leq S_D(i) \leq 2^{2n}-1$, and each
operand has $n$ bits, so $0 \leq A_D(i)\cdot B_D(i) \leq (2^n-1)^2 < 2^{2n}$.
Hence $|\delta| < 2^{2n}$. If $\val{D}{i}(P_D) = 0$, then $2^{2n}$
divides~$\delta$, and therefore $\delta = 0$.
\qed\end{proof}

In practice, obligation (O3) is established by a PAC checker, and (O5) holds by
Lemma~\ref{lem:lift}. The remaining obligations (O1), (O2) and (O4) are the
\emph{interfaces} between the PAC certificate and the circuit. None of
them follows from checking the PAC certificate in isolation. 
For (O1), $C$ has to be obtained from the input
file independently of the tool that generated the PAC certificate. Additionally,
we need to show that $C$ and $D$ are equivalent. 
(O2) holds by Lemma~\ref{lem:circax}
for every axiom in~$\mathrm{Ax}(D)$, but any other axiom requires a separate
argument. Likewise, (O4) holds if $t(\Pi)$ equals~$P_D$ modulo $x^2 = x$,
but any other target, such as the linearized specification of \guessprove,
requires a separate argument. 

\subsection{Certification using Computer Algebra only}
\label{sec:discharge}

In purely algebraic verification, as in \amulettwo when no adder is
substituted, the PAC certificate is generated for the input circuit
itself, that is, $D = C$. In this case, (O1) is established by parsing the
input file in Lean (Sect.~\ref{sec:transcribe}), and (O2) and (O4) reduce to
comparisons with data that Lean computes from the parsed circuit.

\begin{theorem}[Algebraic certification]
\label{thm:alg}
Let $C$ be the circuit parsed from the input file, and let $C$ be well
formed. Let $\Pi$ be a valid PAC certificate such that
$A(\Pi) \subseteq \mathrm{Ax}(C)$ and $t(\Pi)$ equals $P_C$ modulo
$x^2 = x$. Then $C$ multiplies.
\end{theorem}

\begin{proof}
We apply Proposition~\ref{prop:decomp} with $D = C$. (O1) holds trivially,
(O2) follows from $A(\Pi) \subseteq \mathrm{Ax}(C)$ and
Lemma~\ref{lem:circax}, (O3) and (O4) hold by assumption, and (O5) follows
from Lemma~\ref{lem:lift}.
\qed\end{proof}

All hypotheses of Theorem~\ref{thm:alg} are decidable by inspecting $C$
and~$\Pi$. We establish all five obligations in Lean as follows:
\begin{description}
\item[(O1)] The circuit $C$ is defined in Lean by parsing the input file
  (Sect.~\ref{sec:transcribe}) and is not supplied by \amulettwo.
\item[(O2)] Lean computes $\mathrm{Ax}(C)$ from the gates of~$C$ and
  compares it with the axioms of~$\Pi$.
\item[(O3)] Our reflective PAC checker validates every step
  of~$\Pi$ (Sect.~\ref{sec:reflect}).
\item[(O4)] Lean computes $P_C$ from the operand and output literals of~$C$,
  normalizes it and~$t(\Pi)$, and compares the results.
\item[(O5)] Lean checks that $C$ is well formed, so that
  Lemma~\ref{lem:lift} applies.
\end{description}
Thus, even for purely algebraic certificates, we establish more than existing
PAC checkers by linking the certificate to the circuit in the input file. 
\subsection{Certification using Computer Algebra and SAT}
\label{sec:twoplaces}

In both SAT-assisted methods,  \addersub and \guessprove, one interface can no longer be established by
comparing data, since it depends on the function computed by a circuit. We
establish it by SAT refutations and call the facts they provide
\emph{SAT-justified}. 

\paragraph{\Addersub concerning the circuit (O1).} \amulettwo replaces the
final-stage adder of~$C$ by a ripple-carry adder and generates the
certificate for the rewritten circuit~$\Cr$. Hence $D = \Cr$, and
(O2)--(O5) are established as in Sect.~\ref{sec:discharge} with $\Cr$ in
place of~$C$. Since $C$ and~$\Cr$ have the same primary inputs and the
same operand layout, (O1) reduces to the following SAT-justified fact, where
$o_j$ and $o'_j$ denote the $j$-th output literals of $C$ and~$\Cr$:
\[
\text{(F1)}\qquad
\sem{C}{i}(o_j) = \sem{\Cr}{i}(o'_j)
\quad\text{for every input assignment $i$ and every~$j$.}
\]
One such fact is required per design. If \amulettwo finds no adder to
replace, then $D = C$ and no SAT result is required.

\paragraph{\guessprove concerning the axioms (O2).} \talismantwo does not
rewrite the circuit but linearizes the specification (Sect.~\ref{sec:bg}).
Its certificate's circuit~$D$ extends~$C$ by new partial-product gates
$t_{kl}$, and its target is~$P_{\mathrm{TM}}$. They are appended after the gates of~$C$ and defined by
$t_{kl} = a_k \wedge b_l$, so every input assignment determines their values. 
The values of the gates of~$C$ remain unchanged. Hence, $D$ does not restrict the behavior
of~$C$, and (O1) is established without SAT. Since $t_{kl}$ evaluates to
$a_k b_l$, $P_{\mathrm{TM}}$ and~$P_D$ have the same value under every
circuit valuation, which establishes (O4) (Sect.~\ref{sec:gp}).

\talismantwo reduces $P_{\mathrm{TM}}$ by linear relations $p$ between signals
of~$D$. It guesses each such relation~$p$, proves it by SAT, and adds it to
the PAC certificate as an axiom. A relation such as the full-adder
relation $2c + u - x - y - z = 0$ holds because of the joint behavior of
several gates. It is not an initial axiom of
the circuit and cannot be established by comparison with~$\mathrm{Ax}(D)$.
Hence, for each relation~$p$, (O2) therefore requires the
SAT-justified fact
\[
\text{(F2)}\qquad
\val{D}{i}(p) = 0
\quad\text{for every input assignment $i$}.
\]
On our benchmarks, a theorem uses between 62 and 16{,}997 such relations
(Sect.~\ref{sec:evalII}).

\bigskip

The two methods thus employ SAT for different purposes, namely equivalence
checking after a rewrite and the validation of guessed axioms, and its
results enter the proof at different interfaces:
\[
\text{\addersub: SAT} \longrightarrow \text{(O1)},
\qquad
\text{\guessprove: SAT} \longrightarrow \text{(O2)}.
\]

\subsection{From Unsatisfiability to a Fact}
\label{sec:bridge}

A SAT solver establishes that a CNF is unsatisfiable, whereas
Proposition~\ref{prop:decomp} requires facts about circuits such as (F1)
and~(F2). Both facts have the form $\forall i.\ \varphi(i)$, where $i$ ranges over
input assignments and $\varphi(i)$ is the respective equation. To prove such
a fact by SAT, one encodes its negation: a CNF~$F$ describing a
\emph{counterexample}, that is, an input assignment~$i$ for which
$\varphi(i)$ fails. 

For (F1), $F$ is a miter that encodes both circuits and asserts that some
pair of outputs differs (Sect.~\ref{sec:sub}). For (F2), $F$ encodes gates
of~$D$ and asserts that $p \neq 0$ (Sect.~\ref{sec:gp}). If $F$ is
unsatisfiable, no counterexample exists. This argument requires only one direction of the
correctness of the encoding: every counterexample must yield a satisfying
assignment of~$F$, which is obtained by evaluating the encoded circuits
under~$i$. The converse, that every
satisfying assignment corresponds to a counterexample, is not needed for checking a refutation. A fully
verified encoder, such as the bit-blaster of
\texttt{bv\_decide}~\cite{BovingEtAl-BVDecide-OOPSLA25}, proves both
directions. We prove only the first:

\begin{proposition}[Bridge]
\label{prop:bridge}
Let $F$ be a CNF that is satisfiable whenever $\varphi(i)$ fails for some
input assignment~$i$. If the LRAT checker of \texttt{bv\_decide} accepts a
refutation of~$F$, then $\varphi(i)$ holds for every input assignment~$i$.
\end{proposition}

\begin{proof}
By \texttt{verifyCert\_correct} (Sect.~\ref{sec:bg}), $F$ is unsatisfiable,
and hence $\varphi(i)$ cannot fail.
\qed\end{proof}

\noindent
To apply Proposition~\ref{prop:bridge}, Lean computes~$F$ itself with an
encoding function and proves the hypothesis once for this function. The
encoding function takes the same Lean objects as the rest of the proof: for
\addersub, the circuit~$C$ parsed from the input file and the
circuit~$\Cr$ for which the PAC certificate is checked, and for
\guessprove, the gates of the certificate's circuit~$D$ and the guessed
relation~$p$, which is an axiom of the PAC certificate. Hence, the derived fact
refers to exactly the circuits and relations of the final theorem. Only the
encoding function and the proof of its hypothesis are specific to each
method (Lemmas~\ref{lem:circuits-agree} and~\ref{lem:guess} in
Sect.~\ref{sec:instances}), whereas the LRAT checker is used unchanged.

\amulettwo, \talismantwo and CaDiCaL remain untrusted. \amulettwo and
\talismantwo construct their own CNF with a C++ encoder and embed the LRAT
proof~$\pi$ produced by CaDiCaL in the generated Lean file. Lean does not run a SAT solver but
checks~$\pi$ against its own computed~$F$. The two encoders are designed to produce
the same clauses in the same order. If they disagree, the hints in~$\pi$
refer to the wrong clauses, and the checker rejects~$\pi$.

\section{Instantiation of \amulettwo and \talismantwo}
\label{sec:instances}

In this section, we instantiate the scheme of Sect.~\ref{sec:recipe} for
\addersub in \amulettwo and for \guessprove in \talismantwo. Table~\ref{tab:recipe} summarizes
how each method discharges the five obligations.

\begin{table}[t]
\centering
\caption{Discharge of the obligations of Sect.~\ref{sec:relsound} by \addersub (Sect.~\ref{sec:sub}) and \guessprove (Sect.~\ref{sec:gp}).
SAT-justified facts are set in bold. The lower part lists the components that
each method contributes for these facts.}
\label{tab:recipe}
\footnotesize
\begin{tabular*}{\textwidth}{@{\extracolsep{\fill}}lll@{}}
\toprule
 & \Addersub & \guessprove \\
\midrule
Certificate's circuit & $D = \Cr$, adder replaced & $D \supseteq C$, added gates \\
\addlinespace
(O1) circuit & \textbf{(F1)}: $C$ and $\Cr$ agree & prefix comparison \\
(O2) axioms  & Lemma~\ref{lem:circax} & Lemma~\ref{lem:circax} and \textbf{(F2)} \\
(O3) validity & certificate checker & certificate checker \\
(O4) target  & $t(\Pi) = P_{\Cr}$ & $t(\Pi) = P_{\mathrm{TM}}$ \\
(O5) lifting & Lemma~\ref{lem:lift} & Lemma~\ref{lem:lift} \\
\midrule
SAT refutations & one per design & one per used guess \\
CNF             & miter, common part shared & gates $K$ and adders \\
Bridge          & Lemma~\ref{lem:circuits-agree} & Lemma~\ref{lem:guess} \\
Result          & Theorem~\ref{thm:sub} & Theorem~\ref{thm:gp} \\
\bottomrule
\end{tabular*}
\end{table}

\subsection{Adder Substitution in \amulettwo}
\label{sec:sub}

\amulettwo rewrites $C$ into $\Cr$ by replacing the final-stage adder with a
ripple-carry adder and generates a valid PAC certificate~$\Pi$ with
$A(\Pi) \subseteq \mathrm{Ax}(\Cr)$ and target~$P_{\Cr}$. It remains to establish the
SAT-justified fact (F1), that is, that $C$ and~$\Cr$ have the same outputs.


The original check of \amulettwo is a miter that treats the two final-stage
adders as standalone circuits.
It shows that the adders agree on all inputs, but not that the remainder
of~$\Cr$ coincides with the remainder of~$C$ and that the new adder is
connected to it in the same way as the old one, which (F1) requires as well.
We therefore use
a miter of the complete circuits in which the common part is shared.
\label{sec:subbridge}\label{sec:sharedvars}

Every gate of~$\Cr$ outside the new adder is a copy of a gate of~$C$. \amulettwo
records an \emph{alignment}~$\alpha$ that maps each copied gate to its
original. Lean does not trust~$\alpha$ but checks that it is
\emph{faithful}: every aligned gate $g' = \ell'_1 \wedge \ell'_2$ is mapped
to a gate of~$C$ with inputs $\alpha(\ell'_1)$ and~$\alpha(\ell'_2)$, where
$\alpha$ is the identity on primary inputs and constants. By induction along
the gates of~$\Cr$, an aligned gate then has the same value as its original
under every input assignment.

In the CNF $F(C,\Cr,\alpha)$, every aligned gate shares the variable of its
original. Only the gates of~$\Cr$ without alignment, that is, the new adder,
and the gates of~$C$ that are the original of no aligned gate (= the old
replaced adder) are translated to clauses. 
For each output~$j$, a flag~$d_j$ implies that the $j$-th outputs of
the two circuits differ, and a final clause requires one of the flags to be
true. The common part of the two circuits thus does not occur in the CNF;
what remains are the two adders, connected at the shared signals. Since these
signals are unconstrained, a refutation shows that the outputs agree for all
of their values, in particular for the values computed by~$C$. 

\begin{lemma}[Equivalence from a refutation]
\label{lem:circuits-agree}
Let $C$ and $\Cr$ be well formed with the same primary inputs, let $\alpha$
be a faithful alignment, and let the LRAT checker accept a refutation of
$F(C,\Cr,\alpha)$. Then (F1) holds.
\end{lemma}

\begin{proof}
By Proposition~\ref{prop:bridge}, it suffices to show that
$F(C,\Cr,\alpha)$ is satisfiable whenever (F1) fails. Suppose that output~$j$
differs under the input assignment~$i$. We assign to every variable its value
under~$i$ and to each flag~$d_k$ the truth value of the statement that
output~$k$ differs. Since aligned gates have the same values as their
originals, the clauses of both circuits are satisfied. Each flag
$d_k$ is true only if output~$k$ differs, which satisfies the clauses of the
flags, and the final clause is satisfied because $d_j$ is true.
\qed\end{proof}

\begin{theorem}[\Addersub]
\label{thm:sub}
Let $C$ and $\Cr$ be well formed, with the same primary inputs and the same
number of outputs. Let $\Pi$ be a valid PAC certificate with
$A(\Pi) \subseteq \mathrm{Ax}(\Cr)$ such that $t(\Pi)$ equals~$P_{\Cr}$
modulo $x^2 = x$, let $\alpha$ be a faithful alignment, and let the LRAT checker accept a
refutation of $F(C,\Cr,\alpha)$. Then $C$ multiplies.
\end{theorem}

\begin{proof}
We apply Proposition~\ref{prop:decomp} with $D = \Cr$. Obligations
(O2)--(O5) hold as in the proof of Theorem~\ref{thm:alg}. For (O1), the
operand bits are primary inputs and therefore have the same values in both
circuits, and the output bits agree by Lemma~\ref{lem:circuits-agree}.
\qed\end{proof}

\noindent
Hence, one refutation per design suffices, and the size of the miter depends only on the
two adders. 

\subsection{Guess-and-Prove in \talismantwo}
\label{sec:gp}

For \talismantwo, Lean establishes the three interfaces of
Sect.~\ref{sec:twoplaces} as follows.

\paragraph{Circuit (O1).}\label{sec:talisman-transcribe}
Since $D$ contains gates that do not occur in the input file, the proof obligation cannot be
obtained by simply parsing the file. Lean therefore parses~$C$ from the file and
checks that $D$ agrees with~$C$ on its first $m$ gates, where $m$ is the
number of gates of~$C$, and that both circuits have the same inputs, outputs
and operand layout. Since every gate is evaluated from earlier gates and
inputs, the first $m$ gates, and hence the outputs, have the same values in
both circuits. 
For  the remaining gates of~$D$, Lean checks that these are new variables defined from
earlier ones, so they do not restrict the input assignments.

\paragraph{Axioms (O2).}\label{sec:gpbridge}
The axioms of the PAC certificate are the modulus, gate polynomials of~$D$,
and guessed relations. The modulus and the gate polynomials vanish by
Lemma~\ref{lem:circax}, so it remains to establish (F2) for every guessed
relation. A guessed relation is a linear polynomial~$p$ over the variables
of~$D$, such as the full-adder relation of Sect.~\ref{sec:twoplaces}, and
typically follows from a few gates. \talismantwo therefore proves~$p$ using
only the Tseitin encoding of a small set~$K$ of gates of~$D$. The CNF
$G(K,p)$ consists of these clauses and of clauses encoding $p \neq 0$.

\begin{lemma}[Relation from a refutation]
\label{lem:guess}
Let $D$ be well formed, let $K$ be a set of gates of~$D$, and let $p$ be
linear in the variables of~$D$. If the LRAT checker accepts a refutation
of $G(K,p)$, then (F2) holds for~$p$.
\end{lemma}

\begin{proof}
By Proposition~\ref{prop:bridge}, it suffices to show that $G(K,p)$ is
satisfiable if $\val{D}{i}(p) \neq 0$ for some input assignment~$i$. The
values of the variables of~$D$ under~$i$ satisfy the clauses of the Tseitin encoding of~$K$,
and since the integer value of~$p$ is nonzero, they extend to a satisfying
assignment of the clauses encoding $p \neq 0$.
\qed\end{proof}

\noindent
Lean checks for every guessed relation~$p$ that $K$ consists of gates of~$D$, so
that the lemma applies.

\paragraph{Target (O4).}
As a polynomial, $P_{\mathrm{TM}}$ differs from~$P_D$, since every $t_{kl}$ is a
fresh variable. Lean checks that every $t_{kl}$ is a gate of~$D$ with
inputs $a_k$ and~$b_l$. Its value under~$\val{D}{i}$ is then the product of
the values of $a_k$ and~$b_l$, so the double sum in~$P_{\mathrm{TM}}$
evaluates to $A\cdot B$, and $P_{\mathrm{TM}}$ and~$P_D$ have the same value
under every circuit valuation. Hence, (O4) holds if $t(\Pi)$ equals
$P_{\mathrm{TM}}$ modulo $x^2 = x$.

\begin{theorem}[\guessprove]
\label{thm:gp}
Let $C$ be the circuit parsed from the input file, and let $D$ be a
well-formed circuit that agrees with~$C$ on its first $m$ gates, has the same
inputs, outputs and operand layout as~$C$, and contains every $t_{kl}$ as a
gate with inputs $a_k$ and~$b_l$. Let $\Pi$ be a valid PAC
certificate such that $t(\Pi)$ equals $P_{\mathrm{TM}}$ modulo $x^2 = x$ and
every axiom of~$\Pi$ is the modulus, a gate polynomial of~$D$,
or a guessed relation~$p$ that satisfies the hypotheses of
Lemma~\ref{lem:guess} for some set~$K$. Then $C$ multiplies.
\end{theorem}

\begin{proof}
We apply Proposition~\ref{prop:decomp}. (O1) holds since $D$ agrees with~$C$
on its first $m$ gates, (O2) follows from Lemmas~\ref{lem:circax}
and~\ref{lem:guess}, (O3) holds by assumption, (O4) holds since the $t_{kl}$
are the partial products, and (O5) follows from Lemma~\ref{lem:lift}.
\qed\end{proof}

\section{Implementation in Lean}
\label{sec:impl}
\label{sec:onecheck}

All statements of Sects.~\ref{sec:recipe} and~\ref{sec:instances} are proved
in Lean once, for arbitrary circuits, certificates and refutations.
Table~\ref{tab:lean} lists the corresponding Lean theorems, grouped into
those shared by both SAT-assisted methods and those specific to one of them.
Propositions~\ref{prop:decomp} and~\ref{prop:bridge} have no counterpart of
their own, since they only appear as steps in these proofs.

\begin{table}[t]
\centering
\caption{Lean theorems for the statements of Sects.~\ref{sec:recipe}
and~\ref{sec:instances}.}
\label{tab:lean}
\scriptsize
\setlength{\tabcolsep}{3pt}
\begin{tabular*}{\textwidth}{@{\extracolsep{\fill}}l@{\hspace{2pt}}ll@{\hspace{2pt}}ll@{\hspace{2pt}}l@{}}
\toprule
\multicolumn{2}{@{}l}{\emph{shared}} &
\multicolumn{2}{l}{\addersub} &
\multicolumn{2}{l@{}}{\guessprove} \\
\cmidrule(r){1-2}\cmidrule(lr){3-4}\cmidrule(l){5-6}
Thm.~\ref{thm:relsound} & \texttt{checkCert\_sound}
  & Lem.~\ref{lem:circuits-agree} & \texttt{circuits\_agree}
  & (O1) & \texttt{denote\_prefix\_agree} \\
Lem.~\ref{lem:circax} & \texttt{deriveAxioms\_vanish}
  & Thm.~\ref{thm:sub} & \texttt{is\_multiplier\_sub}
  & Lem.~\ref{lem:guess} & \texttt{guessRecordOkTM\_vanish} \\
Lem.~\ref{lem:lift} & in \texttt{is\_multiplier\_alg}
  & & & (O2) & \texttt{axiomsOkTM\_sound} \\
Thm.~\ref{thm:alg} & \texttt{is\_multiplier\_alg}
  & & & (O4) & \texttt{evalPoly\_specPolyTM} \\
& & & & Thm.~\ref{thm:gp} & \texttt{is\_multiplier\_gp} \\
\bottomrule
\end{tabular*}
\end{table}

The shared part comprises the soundness of the PAC checker
(\texttt{checkCert\_sound}, Sect.~\ref{sec:reflect}), the vanishing of the
circuit axioms (\texttt{deriveAxioms\_vanish}), and the lifting to integers
and purely algebraic verification (both in \texttt{is\_multiplier\_alg}). Both
SAT-assisted methods use it without modification and add only what their
interface requires. \Addersub adds the bridge for~(F1)
(\texttt{circuits\_agree}) and its final theorem
(\texttt{is\_multiplier\_sub}), since only (O1) differs from purely
algebraic verification. \guessprove adds the bridge for~(F2)
(\texttt{guessRecordOkTM\_vanish}) and its use for all axioms
(\texttt{axiomsOkTM\_sound}), but also the arguments for (O1)
(\texttt{denote\_prefix\_agree}) and (O4) (\texttt{evalPoly\_specPolyTM}),
since its certificate's circuit and its target differ from those of the
input circuit, and its final theorem (\texttt{is\_multiplier\_gp}).
The final theorems of all three methods have the same conclusion, namely that
the circuit parsed from the input file multiplies.

For a concrete multiplier, \amulettwo or \talismantwo generates a Lean file
that embeds the bytes of the input file as a hexadecimal string
\texttt{fileHex}, defines the circuit as
\texttt{original := fromFile fileHex aBits bBits}, where \texttt{aBits}
and \texttt{bBits} give the operand layout (Sect.~\ref{sec:transcribe}), and
states that \texttt{original} multiplies. 
The proof is a single tactic
call, \texttt{amulet} or \texttt{talisman}, whose arguments are string
literals containing the PAC certificate and,
depending on the method, the rewritten circuit, its alignment and the
refutation, or the circuit~$D$ and the guessed relations with their
refutations (Sect.~\ref{sec:tool}). The tactic applies the final theorem of
the method and discharges all of its hypotheses by evaluating one Boolean
function with \texttt{native\_decide}. These data come from the untrusted
tools \amulettwo, \talismantwo and CaDiCaL, but Lean does not rely on them:
the Boolean function checks them, and
if any of them is wrong, such as an invalid step of the certificate or a
wrong alignment, it returns \texttt{false}, \texttt{native\_decide} fails,
and no theorem is proved.

\subsection{An AIGER parser}
\label{sec:transcribe}

The final theorem stating that the circuit is a correct multiplier 
is meaningful only if it concerns the circuit in the input
file, which is what (O1) requires. If the circuit were supplied by the C++
code of \amulettwo or \talismantwo, an incorrect transcription would make
Lean prove a correct theorem about a different circuit. Certificates checked
with \pacheck have this weakness, since the gate polynomials against which
it checks are written by \amulettwo. We therefore parse the input file in
Lean. The function \texttt{fromFile} reads the header, the output literals
and the AND gates of a binary AIGER file and returns a value of
type \texttt{Aig}. Since the generated file defines the circuit as the result
of \texttt{fromFile} on the embedded bytes of the input file, the final
theorem refers to the circuit described by the file itself.

The only information that \amulettwo and \talismantwo supply about this circuit is the operand
layout, that is, which inputs are the bits $a_k$ and~$b_k$ of
Definition~\ref{def:mult}, since the AIGER format does not necessarily record it (unless added as a comment, which we do not assume).
The layout appears in the generated file as \texttt{aBits} and \texttt{bBits}.
Together with \texttt{fromFile}, it determines what the final theorem
states, and both therefore belong to the trusted specification.

\subsection{A reflective PAC checker}
\label{sec:reflect}\label{sec:lrat}

In our first attempt, we applied a direct approach, which translates each step of the certificate into a Lean
theorem that states the polynomial identity of the step and proves it with a
ring tactic. However, this does not scale, as checking an 8-bit
certificate with 607 step theorems takes already 80\,s. 

We therefore embed the certificate as a single string literal, whose
elaboration does not depend on the number of steps. The Boolean function
\texttt{checkCert} parses the string, normalizes the polynomials, replays
each step and compares the last conclusion with the target.
Theorem~\ref{thm:relsound} is proved once as \texttt{checkCert\_sound} for
every certificate~$\mathit{cert}$ with
$\texttt{checkCert}\ \mathit{cert} = \texttt{true}$. For a concrete
certificate, this hypothesis is established by evaluating \texttt{checkCert}
with \texttt{native\_decide}. 
LRAT refutations are checked in the same way
by \texttt{verifyCert} of \texttt{bv\_decide}~\cite{BovingEtAl-BVDecide-OOPSLA25}, the counterpart of
\texttt{checkCert}. 

Like the certificate, the circuit is again a Lean value, of type
\texttt{Aig}, and the function \texttt{denote} computes the value
$\sem{C}{i}(\ell)$ of a literal~$\ell$ under an input assignment~$i$
(Sect.~\ref{sec:bg}). Since \texttt{denote} is total, it also evaluates
circuits that are not well formed. Well-formedness is therefore checked
separately by the Boolean function \texttt{wf}, and theorems that require
it, such as \texttt{deriveAxioms\_vanish} for Lemma~\ref{lem:circax}, take
$\texttt{wf}\ C = \texttt{true}$ as a hypothesis, which is again established
by evaluation.

\subsection{Changes to \amulettwo and \talismantwo}
\label{sec:tool}

Apart from extending \amulettwo and \talismantwo with printing the Lean files, 
they required the following changes.

\paragraph{\amulettwo.} The original \amulettwo performed the rewrite and the
algebraic verification in separate runs and calls the SAT solver CaDiCaL externaly.
The calls are connected by intermediate files, so that nothing ensures that all three
concern the same circuit. We link CaDiCaL into \amulettwo as a library, so
that a single run performs the rewrite, the equivalence check with LRAT
output and the generation of the certificate on the same in-memory circuit,
and records the alignment of the rewritten circuit
(Sect.~\ref{sec:sub}). If no complex final-stage adder is found, as in ripple-carry
designs, \amulettwo discards the rewrite step
and verifies~$C$ directly.

\paragraph{\talismantwo.} \talismantwo emits an LPAC certificate over the
integers. 
After the reduction, \talismantwo proves every guessed relation that
the certificate cites with LRAT output: it encodes the gates~$K$ and $p \neq 0$ exactly as Lean does
(Sect.~\ref{sec:gp}), solves the CNF with CaDiCaL, and trims the
refutation by dropping deletions and keeping only the derived clauses on
which the empty clause depends. Relations that were found but that the certificate does not
cite are not certified. Finally, \talismantwo records for every axiom
whether it is a gate polynomial or a guessed relation, since the two cannot
reliably be distinguished by their shape, and for every relation the
gates~$K$ from which Lean recomputes the CNF. Lean checks all of these data,
so neither the records nor the trimming need to be trusted.

\section{Evaluation}
\label{sec:eval}

To evaluate our methods, we use the AOKI benchmarks~\cite{HommaEtAl-IEICE06} 
consisting of 192 different multiplier architectures with 64-bit operands. A multiplier 
design combines one of two partial-product generators, simple (\texttt{sp})
or Booth radix-4 (\texttt{bp}), with one of eight accumulators: array
(\texttt{ar}), Wallace tree (\texttt{wt}), balanced delay tree (\texttt{bd}),
overturned-stairs tree (\texttt{os}), Dadda tree (\texttt{dt}),
(4;2)-compressor tree (\texttt{ct}), (7,3)-counter tree (\texttt{cn}) or
redundant binary addition tree (\texttt{ba}). Its final-stage adder is one of
twelve: ripple carry (\texttt{rc}), carry lookahead (\texttt{cl}),
ripple-block carry lookahead (\texttt{rb}), block carry lookahead
(\texttt{bc}), Ladner-Fischer (\texttt{lf}), Kogge-Stone (\texttt{ks}),
Brent-Kung (\texttt{bk}), Han-Carlson (\texttt{hc}), conditional sum
(\texttt{cn}), carry select (\texttt{cs}), or carry skip with fixed
(\texttt{csf}) or variable (\texttt{csv}) block size. A name such as
\texttt{bp-wt-cl} lists the three components.

We ran all our experiments on a cluster with nodes equipped with
Xeon E5-2620 v4 processors running at 2.10 GHz with a 
10 minutes time out and 64 GB of RAM.

For every
design, we measure four wall-clock times: \emph{verification}, a run of the
tool without certificates; \emph{certification}, a second run that solves the
same problem and writes the Lean file; \emph{checking} of this file by Lean,
including 2\,s for loading our development; and the \emph{total} time from
the input file to the checked theorem, that is, certification and checking.
With both methods, all 192 designs are certified.

\subsection{\Addersub}
\label{sec:evalI}

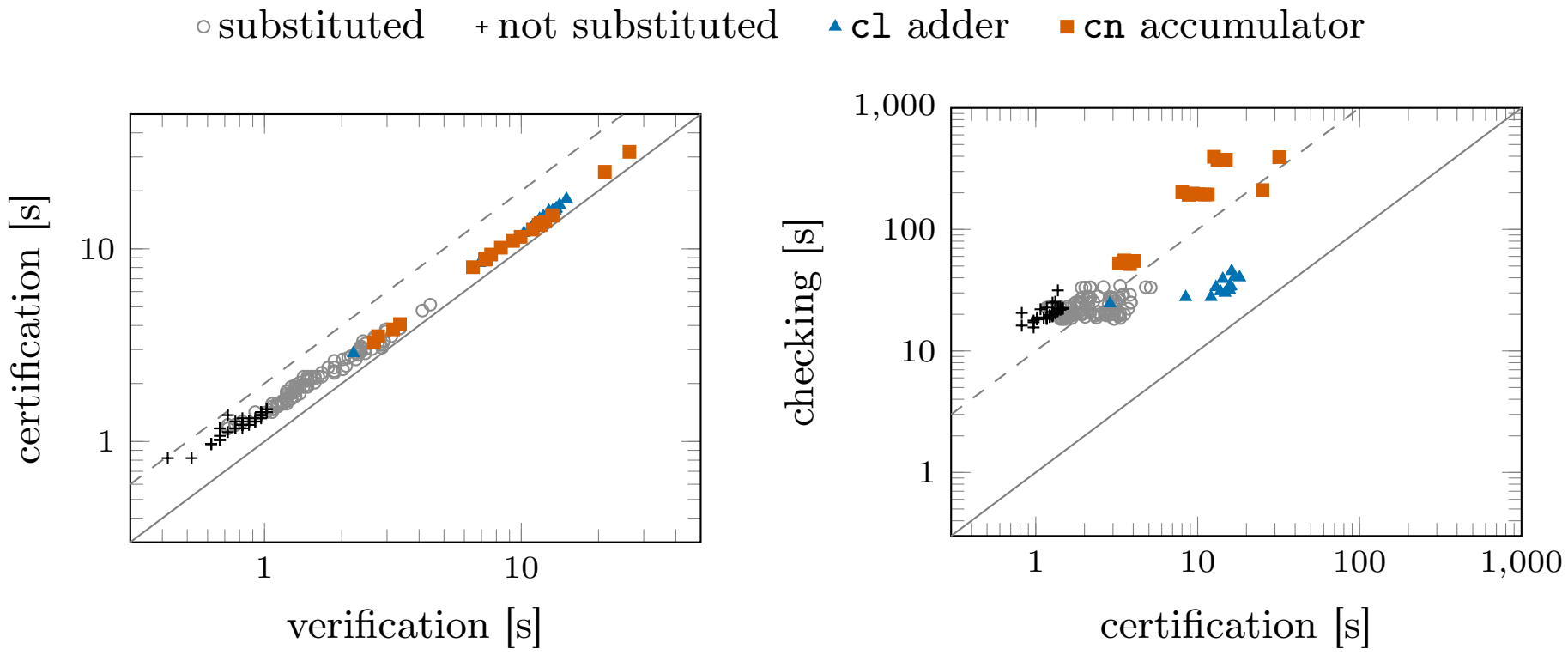
\begin{figure}[t]
\centering
\pgfplotsset{
  evalplot/.style={
    width=0.49\textwidth, height=0.4\textwidth, ylabel near ticks,
    xmin=0.3, xmax=50,
    log ticks with fixed point,
    xlabel={verification [s]},
    tick label style={font=\scriptsize}, label style={font=\footnotesize},
    every axis plot/.append style={only marks, mark size=1.3pt},
    table/col sep=comma, table/x=verify_s,
  },
  ratio/.style={no marks, sharp plot, samples=2, gray, thin},
}
\pgfplotslegendfromname{amuletlegend}\par\smallskip
\begin{tikzpicture}
\begin{loglogaxis}[evalplot, ymin=0.3, ymax=50, ylabel={certification [s]},
  legend to name=amuletlegend, legend columns=4,
  legend style={font=\footnotesize, draw=none, /tikz/every even column/.append style={column sep=1em}}]
\addplot[mark=o, black!45] table[y=certify_s]{data/amulet-sub.csv};
\addlegendentry{substituted}
\addplot[mark=+, black] table[y=certify_s]{data/amulet-nosub.csv};
\addlegendentry{not substituted}
\addplot[mark=triangle*, amcol] table[y=certify_s]{data/amulet-cl.csv};
\addlegendentry{\texttt{cl} adder}
\addplot[mark=square*, tmcol] table[y=certify_s]{data/amulet-cn.csv};
\addlegendentry{\texttt{cn} accumulator}
\addplot[ratio, domain=0.3:50] {x};
\addplot[ratio, dashed, domain=0.3:25] {2*x};
\end{loglogaxis}
\end{tikzpicture}\hfill
\begin{tikzpicture}
\begin{loglogaxis}[evalplot, xmax=1000, ymin=0.3, ymax=1000, table/x=certify_s,
  xlabel={certification [s]}, ylabel={checking [s]}]
\addplot[mark=o, black!45] table[y=check_s]{data/amulet-sub.csv};
\addplot[mark=+, black] table[y=check_s]{data/amulet-nosub.csv};
\addplot[mark=triangle*, amcol] table[y=check_s]{data/amulet-cl.csv};
\addplot[mark=square*, tmcol] table[y=check_s]{data/amulet-cn.csv};
\addplot[ratio, domain=0.3:1000] {x};
\addplot[ratio, dashed, domain=0.3:100] {10*x};
\end{loglogaxis}
\end{tikzpicture}
\caption{\Addersub on the 192 AOKI multipliers: designs with a
(7,3)-counter tree (\texttt{cn}), designs with another accumulator and a
carry-lookahead adder (\texttt{cl}), and the remaining designs with and
without substitution. Solid lines mark the factor 1, dashed lines the
factors 2 (left) and 10 (right).}
\label{fig:amulet}
\end{figure}

\amulettwo substitutes the final-stage adder in 144 designs and verifies~$C$
directly in the remaining 48, whose adders are ripple-carry, conditional-sum
or carry-select adders. Producing the certificates adds little to
verification (Fig.~\ref{fig:amulet}, left). Certification takes 1.07 to 1.95
times as long as verification, with a median factor of 1.34 and a median
difference of 0.5\,s.

Checking the certificates takes 2.0 to 31 times as long as producing them,
with a median of 13 (Fig.~\ref{fig:amulet}, right), and thus dominates the
total time. Certifying all designs takes
at most 337\,MB of RAM for \amulettwo and 5.0\,GB for Lean.

Inspecting the benchmarks in more detail we can make the following observation. 
For the 154 designs with neither a (7,3)-counter tree (\texttt{cn}) nor a carry-lookahead
adder (\texttt{cl}), the Lean file has 8 to 31\,MB, and checking takes 16 to 34\,s.
Carry-lookahead adders lead to LRAT refutations of up to 159\,MB, but the 14
such designs without a (7,3)-counter tree are still checked in 25 to 45\,s,
so even large refutations are cheap to check.
The (7,3)-counter trees
lead to the longest checking times, 52 to 396\,s, which are mostly spent on the PAC
certificate and not the LRAT refutation.

\subsection{\guessprove}
\label{sec:evalII}

\begin{figure}[t]
\centering
\pgfplotsset{
  evalplot/.style={
    width=0.49\textwidth, height=0.4\textwidth, ylabel near ticks,
    xmin=5, xmax=500,
    log ticks with fixed point,
    xlabel={verification [s]},
    tick label style={font=\scriptsize}, label style={font=\footnotesize},
    every axis plot/.append style={only marks, mark size=1.3pt},
    table/col sep=comma, table/x=verify_s,
  },
  ratio/.style={no marks, sharp plot, samples=2, gray, thin},
}
\pgfplotslegendfromname{talismanlegend}\par\smallskip
\begin{tikzpicture}
\begin{loglogaxis}[evalplot, ymin=5, ymax=500, ylabel={certification [s]},
  legend to name=talismanlegend, legend columns=4,
  legend style={font=\footnotesize, draw=none, /tikz/every even column/.append style={column sep=1em}}]
\addplot[mark=o, black!45] table[y=certify_s]{data/talisman-sp.csv};
\addlegendentry{simple (\texttt{sp})}
\addplot[mark=+, black] table[y=certify_s]{data/talisman-bp.csv};
\addlegendentry{Booth (\texttt{bp})}
\addplot[mark=triangle*, amcol] table[y=certify_s]{data/talisman-cl.csv};
\addlegendentry{\texttt{cl} adder}
\addplot[mark=square*, tmcol] table[y=certify_s]{data/talisman-cn.csv};
\addlegendentry{\texttt{cn} accumulator}
\addplot[ratio, domain=5:500] {x};
\addplot[ratio, dashed, domain=5:250] {2*x};
\end{loglogaxis}
\end{tikzpicture}\hfill
\begin{tikzpicture}
\begin{loglogaxis}[evalplot, ymin=5, ymax=500, table/x=certify_s,
  xlabel={certification [s]}, ylabel={checking [s]}]
\addplot[mark=o, black!45] table[y=check_s]{data/talisman-sp.csv};
\addplot[mark=+, black] table[y=check_s]{data/talisman-bp.csv};
\addplot[mark=triangle*, amcol] table[y=check_s]{data/talisman-cl.csv};
\addplot[mark=square*, tmcol] table[y=check_s]{data/talisman-cn.csv};
\addplot[ratio, domain=5:500] {x};
\addplot[ratio, dashed, domain=5:250] {2*x};
\end{loglogaxis}
\end{tikzpicture}
\caption{\guessprove on the 192 AOKI multipliers, grouped as in
Fig.~\ref{fig:amulet}, except that the remaining designs are split by their
partial-product generator. Solid and dashed lines mark the factors 1 and
2.}
\label{fig:talisman}
\end{figure}
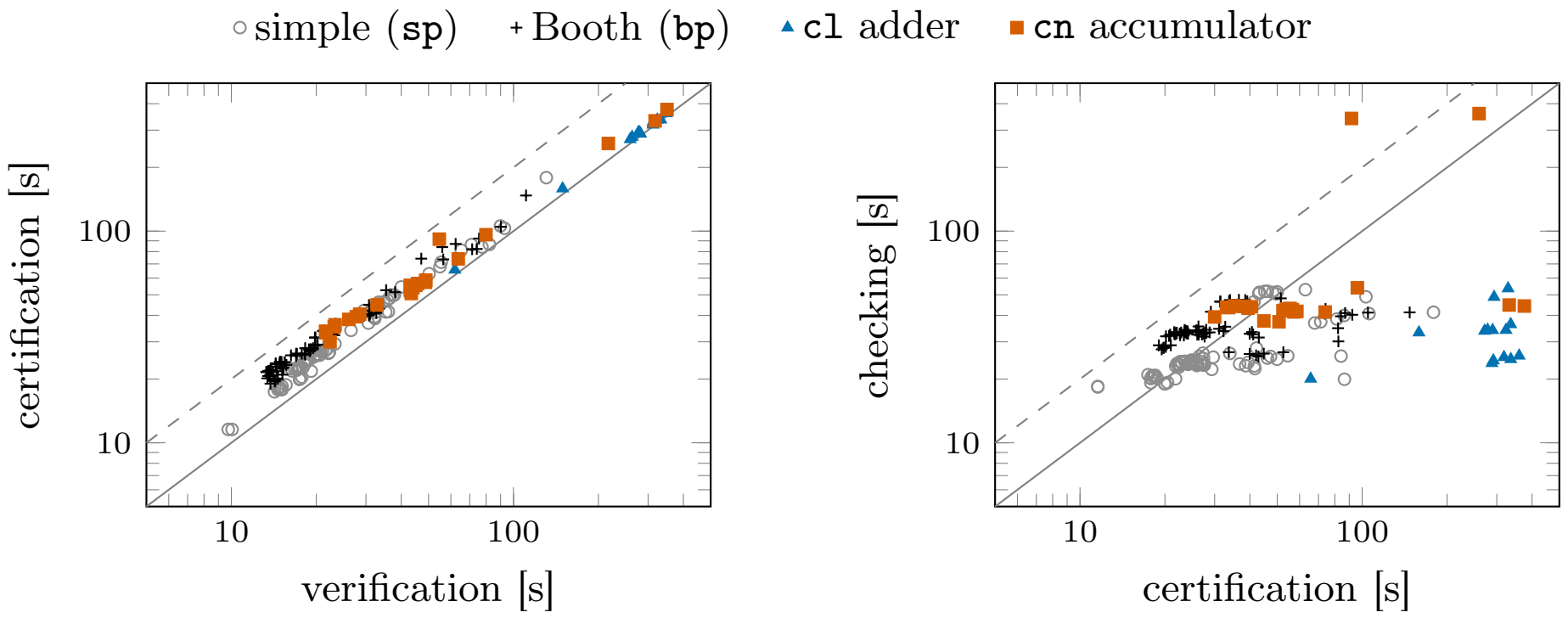

For \talismantwo, certification takes at most 1.7 times as long as
verification, with a median difference of 8.6\,s
(Fig.~\ref{fig:talisman}, left). Unlike for \addersub, checking takes about
as long as certification, with a median ratio of 1.0 and a maximum of 3.7
(Fig.~\ref{fig:talisman}, right). Carry-lookahead adders lead to the longest
verification times, up to 350\,s, since \talismantwo guesses up to
44{,}031 relations for them, but these designs are checked in 20 to 54\,s.
Checking takes a median of 33\,s and more than 54\,s only for
\texttt{sp-cn-cn} and \texttt{sp-cn-cs} (359 and 341\,s), whose PAC
certificates have 45 and 38\,MB, compared to at most 20\,MB for all other
designs. Certifying all designs takes at most 7.3\,GB of RAM for \talismantwo and 4.9\,GB
for Lean.

\section{Trust and Limitations}
\label{sec:trust}

\paragraph{Axioms of the final theorem.} According to \texttt{\#print
axioms}, the final theorem depends on the standard axioms \texttt{propext},
\texttt{Classical.choice} and \texttt{Quot.sound} of Lean and on one
auxiliary axiom introduced by \texttt{native\_decide}, which states that the
combined check of Sect.~\ref{sec:onecheck} evaluates to \texttt{true}. All
other parts of the proof, including the soundness of both checkers and the
statements of Sects.~\ref{sec:recipe} and~\ref{sec:instances}, are checked by
the kernel. The auxiliary axiom is justified by evaluating the check with code
generated by the Lean compiler, so the compiler and the runtime that executes
this code are trusted for this single evaluation. It is in this sense that we call the proof machine-checked.

\paragraph{Kernel reduction.} Evaluating the check by kernel reduction
instead of \texttt{native\_decide} would remove the compiler from the
trusted base. Since the kernel does not unfold definitions by well-founded
recursion, our checker sorts and merges by structural recursion on a fuel
argument, and the variant \texttt{checkCertK} also splits the certificate
into numbers by structural recursion over its characters. With
\texttt{checkCertK}, \texttt{decide +kernel} checks the PAC certificate of a
1-bit multiplier in 6.1\,s. For a 2-bit multiplier
 and for the AOKI multipliers \texttt{sp-ar-rc-4} and
\texttt{sp-ar-rc-8} with 4- and 8-bit operands, it exceeds 40\,GB of memory within about two minutes, whereas
\texttt{native\_decide} checks each of these certificates in at most~0.2\,s.

\paragraph{Untrusted components.} All remaining components are either
checked, or an error in them can only cause a check to fail. This applies to
the rewrite, the SAT and algebraic reasoning, the CNF encoders, the alignment
computed by \amulettwo, the trimming of refutations, and the records and
the multimodular arithmetic of \talismantwo.

\section{Related Work}
\label{sec:related}

\paragraph{SAT-assisted algebraic verification.} \Addersub was introduced by
Kaufmann, Biere and Kauers~\cite{KaufmannBiereKauers-FMCAD19} and implemented
in \amulettwo~\cite{KaufmannBiere-TACAS21,KaufmannBiere-STTT23}. The original
\amulettwo writes the miter to a CNF file for an external SAT solver and
certifies the rewritten circuit by a PAC certificate in a separate run
(Sect.~\ref{sec:tool}). SAT solvers then emitted DRAT proofs, which are checked
by \texttt{drat-trim}~\cite{WetzlerHeuleHunt-SAT14}. 
Translating the SAT proof into PAC yields a
single certificate~\cite{KaufmannBiereKauers-DATE20}, but its axioms, the gate
polynomials of the multiplier, the adder and the miter, are still provided by
the tool rather than derived from the input file.

\guessprove~\cite{HofstadlerKaufmann-CP25} guesses linear relations between
the gates of a subcircuit from sampled assignments and proves each guess by a
SAT call on the subcircuit. \talismantwo~\cite{HofstadlerKaufmannChen-2026,TalismanTool}
performs these steps modulo several primes in parallel but did not log proofs,
which~\cite{HofstadlerKaufmannChen-2026} names as future work.
Temel, Slobodov\'a and Hunt~\cite{TemelSlobodovaHunt-CAV20} verify multipliers
by term rewriting that is proved sound in ACL2 and resort to a SAT solver when
rewriting gets stuck; we are not aware that the SAT result is checked within
ACL2.

\paragraph{Purely algebraic verification.} Konrad and
Scholl~\cite{KonradScholl-FMCAD24,KonradScholl-FMSD26} verify multipliers
without SAT by optimizing the substitution order and the phases of signals.
Their tool emits PAC certificates for \pacheck of up to 365\,GB for 64-bit
multipliers. Kaufmann and Berthomieu~\cite{KaufmannBerthomieu-TACAS25} extract linear
relations from local Gr\"obner bases instead of guessing them, but did not support proof logging.

\paragraph{Verified algebraic proof checkers.} The PAC checker
\pasteque~\cite{KaufmannFleuryBiere-FMCAD20} is verified in Isabelle/HOL. \pasteque is a standalone program generated from
verified code; it returns a verdict rather than a theorem that can be combined
with other results in a proof assistant. In contrast, \texttt{checkCert} runs
within Lean, and its result is a Lean theorem to which the SAT results can be
connected.

\paragraph{Verified SAT proof checkers.} Verified checkers for DRAT and LRAT
proofs exist in CakeML~\cite{TanHeuleMyreen-CakeLPR-TACAS21},
Isabelle/HOL~\cite{Lammich-GRAT-JAR20}, ACL2 and
Coq~\cite{CruzFilipeEtAl-RAT-CADE17}. We reuse the checker of Lean's
\texttt{bv\_decide}~\cite{BovingEtAl-BVDecide-OOPSLA25}. Szeider's
LRAT-Catcher~\cite{Szeider-LRATCatcher-2026} already uses this checker
outside \texttt{bv\_decide} to check DIMACS formulas, so we do not claim this
use as novel. None of these works connects a refutation to a PAC
certificate.

\section{Conclusion}
\label{sec:conclusion}

A PAC certificate is sound only relative to its axioms and to the
circuit they describe. When SAT solving assists algebraic reasoning,
certification must therefore establish not only the validity of each
certificate, but also that the SAT result provides precisely what the
algebraic proof assumes. We have made this connection explicit for two
approaches to combining SAT solving with algebraic multiplier
verification. 
The proof is structured by five obligations. In \addersub,
SAT enters through the certificate's circuit, and in \guessprove through its
axioms. A single bridge, which reuses two verified checkers without
modification, handles both cases. Neither the obligations nor the bridge
depend on the PAC format: any algebraic proof format whose certificates are
sound relative to their axioms, such as Nullstellensatz
certificates~\cite{KaufmannFleuryBiereKauers-PAC-FMSD22}, could take its
place. Investigating this instantiation is part of future work.

Each method now yields a single Lean theorem about the circuit in the input
file. Besides the specification, which includes the AIGER parser and the
operand layout, this theorem relies only on the Lean compiler for a single
evaluation. Both methods certify all 192 AOKI multipliers with 64-bit
operands, and producing the certificates takes at most twice as long as
verification. Checking them takes a median of 13 times as long as producing
them for \addersub, but about as long for \guessprove.

In future work, we plan to certify algebraic divider verification in the
same way. There, a SAT solver proves that certain signals are equivalent or
antivalent under the input constraint of the
divider~\cite{SchollKonrad-DAC20,SchollEtAl-DATE21}, and backward rewriting
uses these facts to simplify polynomials, as \guessprove uses its guessed
relations. Since such facts hold only for inputs that satisfy the constraint,
the obligations have to be extended by this constraint.

\section*{Data Availability Statement}

The Lean development, the modified versions of \amulettwo and \talismantwo, the
generated Lean files, and the scripts that reproduce every table in this
paper are available at \url{10.5281/zenodo.24584511}.

\bibliographystyle{splncs04}
\bibliography{refs}

\end{document}